\documentclass[aps,pre,preprint,groupedaddress]{revtex4-1}
\usepackage{multirow}
\usepackage{graphicx}

\begin{document}

\title{Characterizing and Predicting the Robustness of Power-law Networks}

\author{Sarah LaRocca}
\email[]{larocca@jhu.edu}
\affiliation{Department of Geography and Environmental Engineering, Johns Hopkins University, Baltimore, MD 21218}

\author{Seth Guikema}
\email[]{sguikema@jhu.edu}
\affiliation{Department of Geography and Environmental Engineering, Johns Hopkins University, Baltimore, MD 21218}

\date{\today}


\begin{abstract}


Power-law networks such as the Internet, terrorist cells, species relationships, and cellular metabolic interactions are susceptible to node failures, yet maintaining network connectivity is essential for network functionality. Disconnection of the network leads to fragmentation and, in some cases, collapse of the underlying system. However, the influences of the topology of networks on their ability to withstand node failures are poorly understood. Based on a study of the response of 2,000 power-law networks to node failures, we find that networks with higher nodal degree and clustering coefficient, lower betweenness centrality, and lower variability in path length and clustering coefficient maintain their cohesion better during such events. We also find that network robustness, i.e., the ability to withstand node failures, can be accurately predicted a priori for power-law networks across many fields. These results provide a basis for designing new, more robust networks, improving the robustness of existing networks such as the Internet and cellular metabolic pathways, and efficiently degrading networks such as terrorist cells.

\end{abstract}



\maketitle



Being able to quickly and efficiently estimate the ability of a given network to withstand node failures, that is, its robustness, is central to being able to manage critical networks in the real world. However, there does not yet exist a method for estimating the robustness of networks quickly and accurately based on the topological characteristics of the network, and the existing understanding of the influence of topological characteristics on network robustness if limited. In this paper we focus on scale-free networks and develop such a model for use not only in small cases of failure events but also in modeling large-scale failure events induced by common-cause failures such as natural disasters in which large portions of networks fail.

Scale-free networks exhibit a power-law nodal degree distribution where the probability that a given node is connected to $k$ other nodes is described by $P(k) \sim k^{-\gamma}$ \citep{Barabasi1999}. Empirical evidence indicates that nodal degree in many real networks is limited by the physical costs of adding links to a node. Such networks can be described by adding an exponential cutoff to the power-law distribution $P(k) \sim k^{-\gamma}e^{-(k/\kappa)}$, where $\kappa$ is the cutoff above which it becomes physically very costly to add links to a node \citep{Amaral2000,Jeong2000,Newman2001,Clauset2009}. Scale-free networks have been demonstrated to be tolerant to random failures \citep{Albert2000}. However, the combined influence of individual measures of network topology on failure tolerance has not been studied. Without an understanding of the relationship between topology and robustness to node failures, we are limited in our ability to design failure-tolerant networks across many different domains and 
in our ability to efficiently degrade networks that we wish to attack. Here, we present a systematic study of the effects of topological characteristics on power-law network fault tolerance, and we develop a topology-based statistical approach for estimating the ability of a network to tolerate node failures.



Prior work on network robustness focuses on relatively small numbers of networks due to the limited number of real networks for which data is available \citep{Jeong2001, Holme2002, Albert2004, Crucitti2004, Estrada2006, Montoya2006, Liu2011}. However, this significantly limits the statistical strength of the insights that can be drawn from the analysis. To overcome this limitation, we begin by randomly generating 2,000 networks with degree distributions following a power-law with exponential cutoff and distribution parameters representative of scale-free networks in a variety of domains \citep{Albert2002}. We use five pairs of distribution parameters, based loosely on the network data presented in \citep{Albert2002}: $(\gamma = 1.1, \kappa = 40)$, $(\gamma = 2.0, \kappa = 900)$, $(\gamma = 2.1, \kappa = 400)$, $(\gamma = 2.4, \kappa = 2000)$, $(\gamma = 1.7, \kappa = 200)$.  We generate 400 random networks for each parameter combination: 20 networks for each of 20 sizes, $N = {100; 126; 177; 205; 299; 313; 
336; 367; 387; 482; 513; 540; 557; 592; 621; 758; 821; 936; 967; 1,000}$. The network sizes between 100 and 1000 are generated from a uniform random distribution.

After generating these networks, we calculate the mean, standard deviation, minimum, and maximum values of four topological characteristics for \emph{each} network individually.  Table \ref{tab:networkStats} presents the mean, standard deviation, minimum, and maximum of each of these network-level summary statistics: nodal degree ($\mu = 5.13$), clustering coefficient \citep{Watts1998} ($\mu = 0.289$), betweenness centrality \citep{Freeman1977} ($\mu = 746$) and path length ($\mu = 2.47$).  The ranges for our network characteristics are similar to ranges for real networks such as the Internet, movie actors, scientific paper co-authorship, metabolic reactions, food webs, and word synonyms as presented in \cite{Albert2002}. The mean degree of our networks ranges from 2.3 to 11.2 and the mean degree of networks in \cite{Albert2002} ranges from 2.39 to 173, though only a few non-physical networks such as word associations and social networks reported in \cite{Albert2002} have a mean nodal degree greater than 18. 
The mean clustering coefficient of our networks ranges from 0.067 to 0.61 and the mean clustering coefficient of networks in \cite{Albert2002} ranges from 0.066 to 0.79. The mean path length of our networks ranges from 2.0 to 3.3 and the mean path length of networks in \cite{Albert2002} ranges from 2.4 to 18.7, though the path lengths for power-law networks reported by \cite{Albert2002} are close to those of our network with the exception of several social networks.



\begin{table}
\caption{\label{tab:networkStats}Summary of the topological characteristics of generated networks.}
\begin{ruledtabular}
\begin{tabular}{lccccc}

\textbf{Parameter measure} & \textbf{Within-network deviation} & \textbf{Mean} & \textbf{Std. dev.} & \textbf{Min.} & \textbf{Max.}\\

Network size ($n$) & -- & 505 & 272 & 100 & 1000\\
\multirow{4}{*}{Degree ($k$)} & Mean & 5.1 & 2.2 & 2.3 & 11.2\\
& Minimum & 1.0 & 0.0 & 1.0 & 1.0\\
& Maximum & 307 & 220 & 24 & 989\\
& Standard deviation & 17.7 & 5.7 & 4.5 & 31.7\\
\multirow{4}{*}{Betweenness centrality ($Cb$)} & Mean & 746 & 452 & 105 & 2,278\\
& Minimum & 0.0 & 0.0 & 0.0 & 0.0\\
& Maximum & 192,468 & 229,924 & 1,808 & 992,170\\
& Standard deviation & 8,433 & 7,495 & 316 & 31,375\\
\multirow{4}{*}{Clustering coefficient ($C$)} & Mean & 0.29 & 0.09 & 0.07 & 0.61\\
& Minimum & 0.0 & 0.0 & 0.0 & 0.0\\
& Maximum & 1.0 & 0.0 & 1.0 & 1.0\\
& Standard deviation & 0.39 & 0.07 & 0.13 & 0.48\\
\multirow{4}{*}{Path length ($\ell$)} & Mean & 2.5 & 0.3 & 2.0 & 3.3\\
& Minimum & 1.0 & 0.0 & 1.0 & 1.0\\
& Maximum & 4.7 & 0.96 & 3.0 & 8.0\\
& Standard deviation & 0.5 & 0.1 & 0.1 & 1.1\\

\end{tabular}
\end{ruledtabular}
\end{table}



We repeatedly and independently simulate 100 random node failure scenarios for each network, with equal failure probability for each node, varying the number of nodes failed from $0.10 N_0$ up to $0.75 N_0$, where $N_0$ is the number of nodes in the initial, unperturbed network. These node failure events result in disconnection of one or more nodes from the remainder of the network. Our measure of network robustness is:
\begin{equation}
S^P = \frac{N^p_f}{N_0},
\end{equation} 
where $N^p_f$ is the total number of nodes in the largest connected component after $p$ percent of nodes have failed. $S^P$ thus gives us the relative size of the largest connected component, a measure of resistance of the network to node failure events \citep{Albert2000}. We calculate $S$ for each failure scenario in each network at four failure levels: 10, 25, 50, and 75 percent of nodes failed.



Because it is a ratio of largest connected component after failures to initial network size, our observed data, $S$, in this analysis is restricted to the interval (0,1). Ferrari and Cribari-Neto (\citeyear{Ferrari2004a}) developed an approach for multiple linear regression modeling of such Beta-distributed response data. We use this approach to develop Beta regression models for network robustness. Our initial data set includes sixteen explanatory variables: minimum, maximum, mean, and standard deviation values for each of the four topological characteristics of networks previously described. We removed variables with standard deviation equal to zero from our data set because they will have no impact in a regression model. These variables are minimum degree, minimum betweenness, minimum clustering coefficient, maximum clustering coefficient, and minimum path length. To reduce correlation between the remaining variables, we remove maximum degree, standard deviation of degree, standard deviation of 
betweenness, and mean path length from our data set. We then standardize the remaining variables for easier interpretability of model results.

We fit Beta regression models to our reduced data by performing maximum likelihood estimation using the Betareg package \citep{Cribari2010} in [R] \citep{R2012}. Because we calculate $S$ after four levels of node removals (10, 25, 50, and 75 percent), we develop four separate models, with a $S$ for a single level of node failures (10, 25, 50, and 75 percent) as the response variable in each model.  After fitting a given initial model, we iteratively remove all covariates from the model that are not statistically significant. That is, for a given model, we remove the explanatory variable with the highest p-value, refit the model, and repeat until all variables are statistically significant at the level of $\alpha = 0.05$.

We then test the predictive accuracy of the models with repeated random holdout validation. We randomly split our initial dataset into a training data set (80\% of initial dataset = 1,600 networks) and a validation data set (20\% of initial dataset = 400 networks). We use our training data to fit regression models for $S$ for each level of node removal (10, 25, 50, and 75 percent). We then use these regression models to predict $S$ for each level of node removal for each network in our validation dataset. We also simulate 100 sets of node failure events for each network in the validation dataset.  Finally, we compare the predicted $S$ to the simulated $S$ for each network in the validation dataset. We repeat this process 100 times (beginning with the random split of our initial dataset) for a 100-fold random holdout cross-validation.



We find that five topological characteristics are statistically significant ($p < 0.05$) predictors of network robustness across all levels of node removal: mean nodal degree, mean betweenness centrality, mean clustering coefficient, standard deviation of clustering coefficient, and standard deviation of path length. We also find that in all four cases, incorporating the topological characteristics increases the fit of the model relative to an intercept-only model by a statistically significant amount ($p < 2.2e-16$). Together, these suggest that topological characteristics are associated with network robustness to node failures and thus may be useful predictors of network robustness.

\begin{figure}
\includegraphics{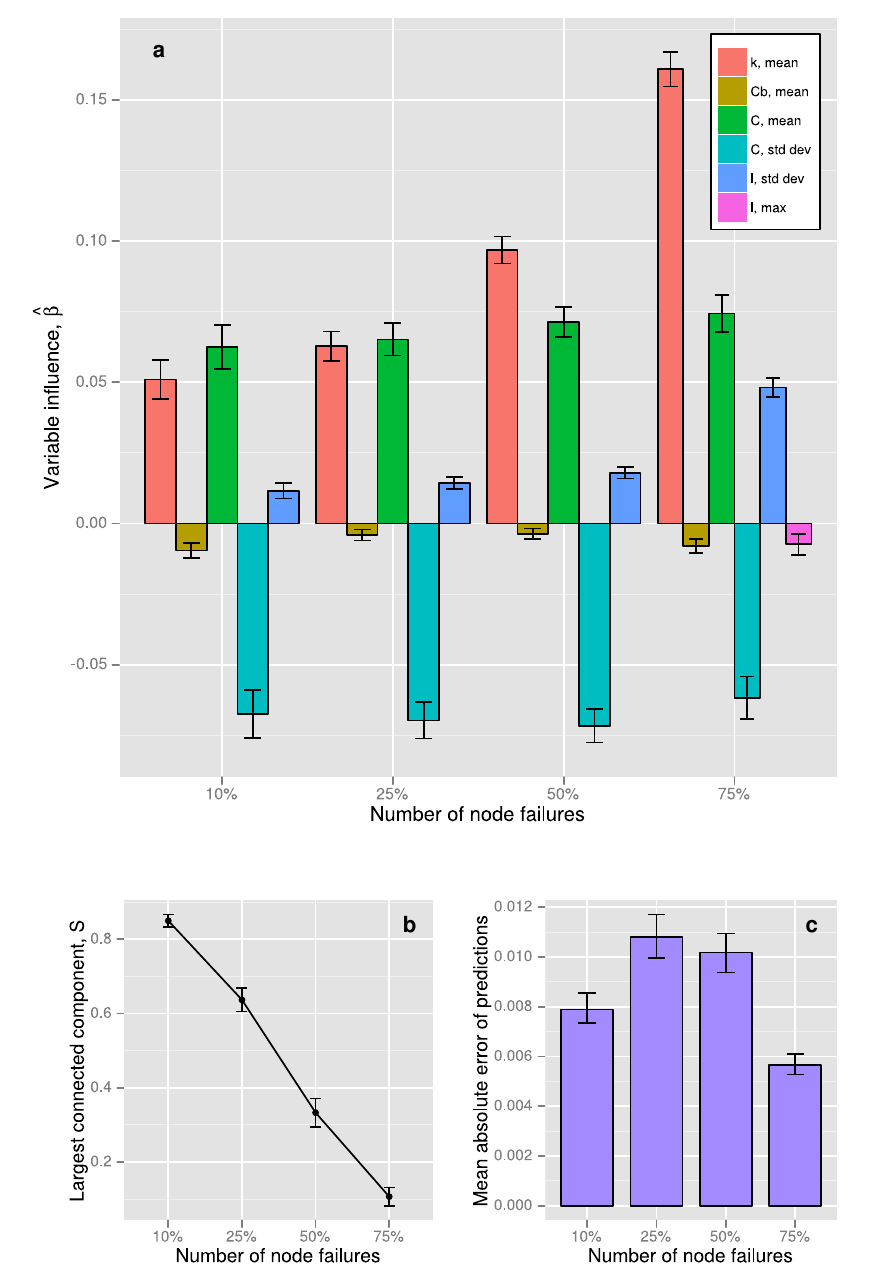}
\caption{\label{fig:model}Beta regression models of network robustness as a function of initial network topology.} 
\end{figure}

Across all ranges of node removal, higher mean nodal degree, mean clustering coefficient, and standard deviation of path length all have positive influences on $S$, while higher mean betweenness centrality and standard deviation of the clustering coefficient have negative influences on $S$. Figure \ref{fig:model}a shows the influence of these initial topological characteristics on network robustness; Figure \ref{fig:model}b shows that mean network robustness, that is, the size of the largest connected component, $S$, decreases as the number of node failures increases as expected; error bars give the standard error of $S$. We also test the predictive ability of our regression models by performing holdout validation on our data as described above. Figure \ref{fig:model}c shows the mean absolute errors of the models' predictions (represented by the purple bars), which are small compared to the true values of $S$ (Figure \ref{fig:model}b). The error bars give the 95\% confidence interval for the prediction 
errors. Overall, our models fit the data well and indicate that topological characteristics are important predictors of network robustness. Our models also yield accurate out of sample predictions of network robustness. We next discuss the influence of the statistically significant topological characteristics in more detail to draw insights into their influences on network robustness.



The mean of the clustering coefficient, $C$, a measure of how locally connected nodes are, is the most important topological characteristic in determining $S$ when small fractions of nodes are removed. For a network of a given degree distribution, higher local connectivity (e.g., clustering) implies that a more locally redundant set of edges exists. When one node is removed, this locally redundant set of edges decreases the chance that nodes will become disconnected from the network, increasing $S$. Our results confirm this. Furthermore, our results show that the influence of $C$ on $S$ remains relatively constant from 10\% of nodes removed through 75\% of nodes removed, though it becomes less influential than nodal degree, $k$ at 50\% and 75\% of nodes removed. Maintaining high local clustering is thus important across a range of magnitudes of impacts to networks, with increased clustering effectively offering local redundancy.

Our model shows that the mean of $k$ is nearly as important as $C$ at low levels of node removal but quickly becomes the most influential topological characteristic. Higher average nodal degree would imply a network in which connections are concentrated in a smaller number of nodes. Random node failures are equally likely to remove any node from the network. As $E[k]$ increases, it becomes more likely that a random node removal would remove a weakly connected node that other nodes do not rely on for their connection to the network. This would decrease the likelihood of decreasing $S$. This effect increases dramatically as the fraction of nodes failed increases. This would be expected because at the higher levels of node removal, the concentration of edge connections into a few central nodes would be even more important in making a network robust against \emph{random} node failures. In our regression model, $\mu$ is the mean of $S$. With our logit link function, we have:

\begin{equation}
\frac{\partial \frac{S}{1-S}}{\partial x_i} = e^{\beta_i}
\end{equation}
This gives a measure of the influence on a given variable, $x_i$ on the ratio of fraction connected to fraction disconnected. At the 10\% node removal level, increasing the mean of $k$ by 1 unit increases the ratio of fraction connected to fraction disconnected by 1.05 on average, whereas at 75\% node removal, a 1 unit increase in the mean of $k$ increases the ratio of fraction connected to fraction disconnected by 1.17. Note that we would expect substantially different results for targeted node failure events where one might expect higher degree nodes to be targeted for removal first.

The standard deviation of path length also exerts a positive influence on $S$ and its influence increases with number of node failures. However, relative to both the mean of $k$ and $C$, its influence is lower. Higher variability in the path lengths implies greater diversity in the nodes traversed. This additional redundancy in paths between nodes should make it less likely that a given node will be disconnected by node failures, all else being equal in the network. At the same time, the maximum shortest path length is statistically significant only when 75\% of the nodes are removed. Longer shortest path lengths require more nodes to be traversed to maintain connectivity, decreasing the opportunities for path redundancy, again, holding all else fixed. This reinforces the insight that diversity in paths traversed is important because it increases path redundancy.

In contrast to the clustering coefficient mean, increasing its standard deviation negatively influences network robustness, and the absolute value of this influence is approximately equal to that of the (positive) influence of the mean of $C$ across all levels of node removal. To understand this influence it is important to note that the mean of the clustering coefficient is relatively small ($2.5$). As the standard deviation of $C$ increases, it becomes more likely that a node randomly selected for failure would have a low value of $C$. Removing this node would have a larger impact on surrounding nodes because it is not as locally connected as nodes with a higher $C$, leading to an increased chance of additional nodes depending on it but not other nodes becoming disconnected.

Betweenness centrality also exhibits a negative influence on $S$, although its influence is less than that of the other variables. Betweenness centrality of a given node quantifies the relative number of shortest paths that will become longer if that node is removed from the graph. Longer shortest paths are then more susceptible to being severed by other node failures, resulting in decreased network robustness.

In addition to providing a fundamental understanding of the relative influence of different topological characteristics on network robustness, our model can also be used to predict the robustness of real-world power-law networks using simple information about the network's initial topology.  We test our regression models' predictive capabilities on three real-world power-law networks: the Ythan estuary food web \citep{Huxham1996} ($k_{mean} = 8.84$, $Cb_{mean} = 187$, $C_{mean} = 0.217$, $\ell_{mean} = 2.41$); the metabolic pathway graph for the bacteria Escherichia coli \citep{Keseler2011} ($k_{mean} = 3.10$, $Cb_{mean} = 806$, $C_{mean} = 0.076$, $\ell_{mean} = 5.43$); and the terrorist network of 9-11 hijackers and their affiliates \citep{Shaikh2007} ($k_{mean} = 5.00$, $Cb_{mean} = 126$, $C_{mean} = 0.472$, $\ell_{mean} = 3.06$). See Supplemental Material for figures showing the topology of each of these networks before and after one set of realizations of random node failures.

\begin{figure}
\includegraphics{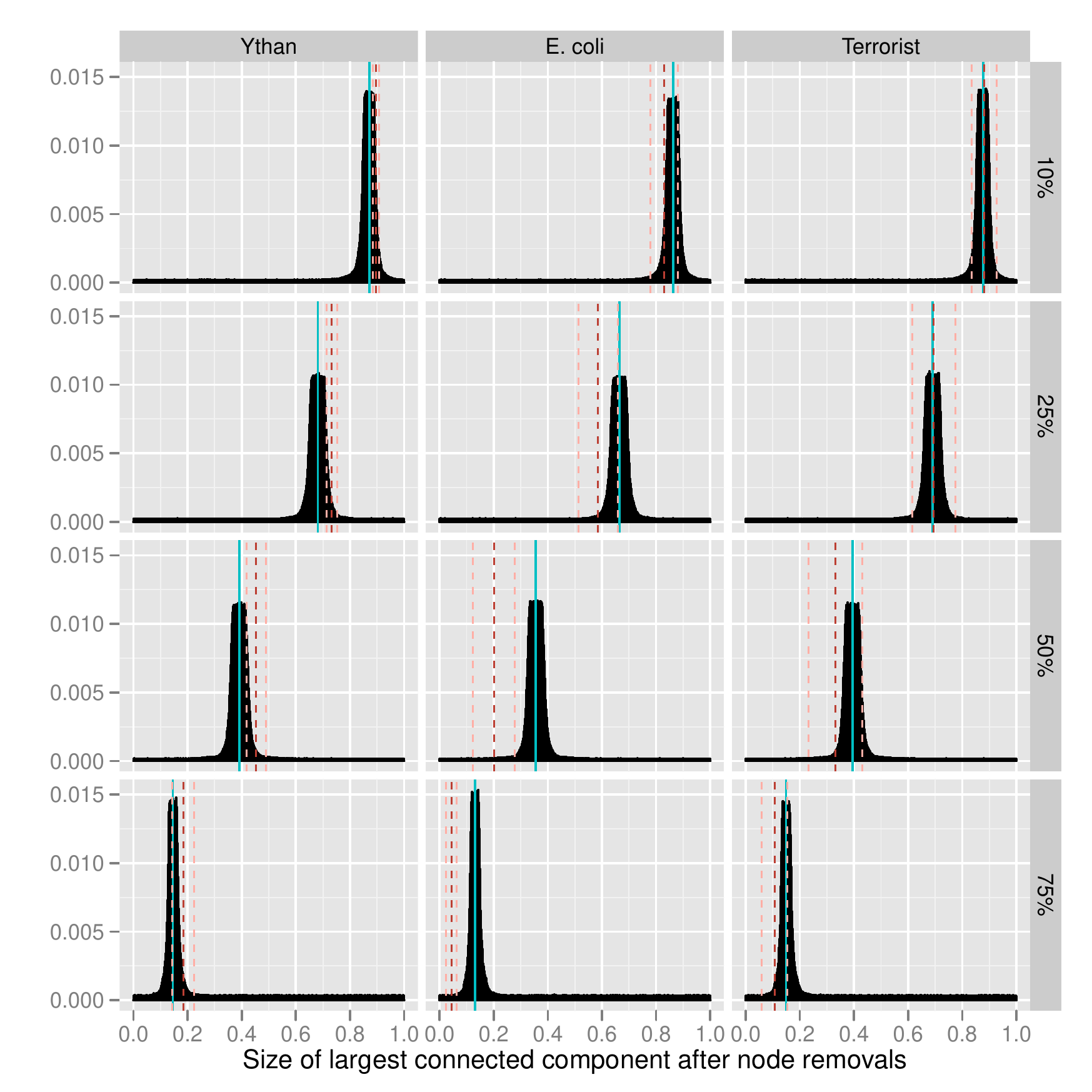}
\caption{\label{fig:predict}Robustness of real-world networks: predictions and true values.}
\end{figure}

We find that we are able to use our statistical approach to estimate $S$ for these real networks with a high level of accuracy, particularly for small fractions of node removals (i.e., 10\% and 25\%) (Figure \ref{fig:predict}). In Figure \ref{fig:predict}, the vertical blue lines show our predictions of network robustness, $S$, for each of the three networks at four levels of node failure. The corresponding histograms give the probability density functions of our simulated $S$ values. The vertical red dashed lines indicate the simulated values of network robustness for each network and failure combination and the vertical pink dashed lines indicated the mean simulated value plus and minus one standard error. The E. coli network was the hardest to predict accurately, with the actual (simulated) $S$ lying outside of the 95\% prediction interval for the two highest levels of node removal. The terrorist network was the easiest to predict accurately, with the 95\% prediction interval continuing the true value for all cases.



In summary, we demonstrate that there is a statistically significant relationship between the initial topological properties of scale-free networks and their corresponding robustness. Our stastical models are generalizable to large-scale, realistic networks and provide strong insights into the effects of topology on robustness. We find that although the relative influence of different topological measures varies depending on the level of network disturbance (e.g., number of nodes removed), the direction of the influence of a given characteristic is always the same.  Specifically, higher nodal degree and clustering coefficient, lower betweenness centrality, and lower variability in path length and clustering coefficient imply greater network robustness.  This improved understanding of the impact of network topology on robustness has many applications and benefits in the context of operations research.  Because our models allow us to rapidly and accurately estimate network robustness, they can be used to 
prioritize improvement efforts among multiple existing networks and to allocate resources to those networks. Additionally, such robustness estimates can be incorporated into the optimization of single networks, both for the design of new networks and for improving (or degrading) existing networks.  We show that using our robustness estimates to identify optimal attack strategies on a terrorist network provides a closer match to the true optimal strategy than basing the attack strategy on nodal degree. Finally, the relative simplicity of our models, both in required data and in computational complexity, makes them a highly practical and efficient tool for aiding real-world decision-making.


\begin{acknowledgments}

This material is based upon work supported by the National Science Foundation under Grant No. CMMI-0968711. Sarah LaRocca is also funded by the National Science Foundation Graduate Research Fellowship under Grant No. DGE-1232825.
\end{acknowledgments}


\bibliography{networks.bib}

\begin{thebibliography}{10}%
\makeatletter
\providecommand \@ifxundefined [1]{%
 \ifx #1\undefined \expandafter \@firstoftwo
 \else \expandafter \@secondoftwo
\fi
}%
\providecommand \@ifnum [1]{%
 \ifnum #1\expandafter \@firstoftwo
 \else \expandafter \@secondoftwo
\fi
}%
\providecommand \enquote [1]{``#1''}%
\providecommand \bibnamefont  [1]{#1}%
\providecommand \bibfnamefont [1]{#1}%
\providecommand \citenamefont [1]{#1}%
\providecommand\href[0]{\@sanitize\@href}%
\providecommand\@href[1]{\endgroup\@@startlink{#1}\endgroup\@@href}%
\providecommand\@@href[1]{#1\@@endlink}%
\providecommand \@sanitize [0]{\begingroup\catcode`\&12\catcode`\#12\relax}%
\@ifxundefined \pdfoutput {\@firstoftwo}{%
 \@ifnum{\z@=\pdfoutput}{\@firstoftwo}{\@secondoftwo}%
}{%
 \providecommand\@@startlink[1]{\leavevmode\special{html:<a href="#1">}}%
 \providecommand\@@endlink[0]{\special{html:</a>}}%
}{%
 \providecommand\@@startlink[1]{%
  \leavevmode
  \pdfstartlink
   attr{/Border[0 0 1 ]/H/I/C[0 1 1]}%
   user{/Subtype/Link/A<</Type/Action/S/URI/URI(#1)>>}%
  \relax
 }%
 \providecommand\@@endlink[0]{\pdfendlink}%
}%
\providecommand \url  [0]{\begingroup\@sanitize \@url }%
\providecommand \@url [1]{\endgroup\@href {#1}{\urlprefix}}%
\providecommand \urlprefix [0]{URL }%
\providecommand \Eprint[0]{\href }%
\@ifxundefined \urlstyle {%
  \providecommand \doi [1]{doi:\discretionary{}{}{}#1}%
}{%
  \providecommand \doi [0]{doi:\discretionary{}{}{}\begingroup
  \urlstyle{rm}\Url }%
}%
\providecommand \doibase [0]{http://dx.doi.org/}%
\providecommand \Doi[1]{\href{\doibase#1}}%
\providecommand \bibAnnote [3]{%
  \BibitemShut{#1}%
  \begin{quotation}\noindent
    \textsc{Key:}\ #2\\\textsc{Annotation:}\ #3%
  \end{quotation}%
}%
\providecommand \bibAnnoteFile [2]{%
  \IfFileExists{#2}{\bibAnnote {#1} {#2} {\input{#2}}}{}%
}%
\providecommand \typeout [0]{\immediate \write \m@ne }%
\providecommand \selectlanguage [0]{\@gobble}%
\providecommand \bibinfo [0]{\@secondoftwo}%
\providecommand \bibfield [0]{\@secondoftwo}%
\providecommand \translation [1]{[#1]}%
\providecommand \BibitemOpen[0]{}%
\providecommand \bibitemStop [0]{}%
\providecommand \bibitemNoStop [0]{.\EOS\space}%
\providecommand \EOS [0]{\spacefactor3000\relax}%
\providecommand \BibitemShut [1]{\csname bibitem#1\endcsname}%
\bibitem{Barabasi1999}%
  \BibitemOpen
  \bibfield{author}{%
  \bibinfo {author} {\bibfnamefont{A.}~\bibnamefont{Barab\'{a}si}}\ and\
  \bibinfo {author} {\bibfnamefont{R.}~\bibnamefont{Albert}},\ }%
  \bibfield{journal}{%
  \bibinfo {journal} {Science}\ }%
  \textbf{\bibinfo {volume} {286}},\ \bibinfo {pages} {509} (\bibinfo {month}
  {Oct.}\ \bibinfo {year} {1999})%
  \bibAnnoteFile{NoStop}{Barabasi1999}%
\bibitem{Amaral2000}%
  \BibitemOpen
  \bibfield{author}{%
  \bibinfo {author} {\bibfnamefont{L.}~\bibnamefont{Amaral}}, \bibinfo {author}
  {\bibfnamefont{A.}~\bibnamefont{Scala}}, \bibinfo {author}
  {\bibfnamefont{M.}~\bibnamefont{Barth\'{e}l\'{e}my}},\ and\ \bibinfo {author}
  {\bibfnamefont{H.}~\bibnamefont{Stanley}},\ }%
  \bibfield{journal}{%
  \bibinfo {journal} {Proceedings of the National Academy of Sciences of the
  United States of America}\ }%
  \textbf{\bibinfo {volume} {97}},\ \bibinfo {pages} {11149} (\bibinfo {year}
  {2000})%
  \bibAnnoteFile{NoStop}{Amaral2000}%
\bibitem{Jeong2000}%
  \BibitemOpen
  \bibfield{author}{%
  \bibinfo {author} {\bibfnamefont{H.}~\bibnamefont{Jeong}}, \bibinfo {author}
  {\bibfnamefont{B.}~\bibnamefont{Tombor}}, \bibinfo {author}
  {\bibfnamefont{R.}~\bibnamefont{Albert}}, \bibinfo {author}
  {\bibfnamefont{Z.~N.}\ \bibnamefont{Oltvai}},\ and\ \bibinfo {author}
  {\bibfnamefont{A.-L.}\ \bibnamefont{Barab\'{a}si}},\ }%
  \bibfield{journal}{%
  \bibinfo {journal} {Nature}\ }%
  \textbf{\bibinfo {volume} {407}},\ \bibinfo {pages} {651} (\bibinfo {month}
  {Oct.}\ \bibinfo {year} {2000})%
  \bibAnnoteFile{NoStop}{Jeong2000}%
\bibitem{Newman2001}%
  \BibitemOpen
  \bibfield{author}{%
  \bibinfo {author} {\bibfnamefont{M.}~\bibnamefont{Newman}},\ }%
  \bibfield{journal}{%
  \bibinfo {journal} {Physical Review E}\ }%
  \textbf{\bibinfo {volume} {64}},\ \bibinfo {pages} {016131} (\bibinfo {month}
  {Jul.}\ \bibinfo {year} {2001}),\ ISSN \bibinfo {issn} {1539-3755}%
  \bibAnnoteFile{NoStop}{Newman2001}%
\bibitem{Clauset2009}%
  \BibitemOpen
  \bibfield{author}{%
  \bibinfo {author} {\bibfnamefont{A.}~\bibnamefont{Clauset}}, \bibinfo
  {author} {\bibfnamefont{C.~R.}\ \bibnamefont{Shalizi}},\ and\ \bibinfo
  {author} {\bibfnamefont{M.}~\bibnamefont{Newman}},\ }%
  \bibfield{journal}{%
  \bibinfo {journal} {SIAM Review}\ }%
  \textbf{\bibinfo {volume} {51}},\ \bibinfo {pages} {661} (\bibinfo {year}
  {2009})%
  \bibAnnoteFile{NoStop}{Clauset2009}%
\bibitem{Albert2000}%
  \BibitemOpen
  \bibfield{author}{%
  \bibinfo {author} {\bibfnamefont{R.}~\bibnamefont{Albert}}, \bibinfo {author}
  {\bibfnamefont{H.}~\bibnamefont{Jeong}},\ and\ \bibinfo {author}
  {\bibfnamefont{A.-L.}\ \bibnamefont{Barab\'{a}si}},\ }%
  \bibfield{journal}{%
  \bibinfo {journal} {Nature}\ }%
  \textbf{\bibinfo {volume} {406}},\ \bibinfo {pages} {378} (\bibinfo {month}
  {Jul.}\ \bibinfo {year} {2000})%
  \bibAnnoteFile{NoStop}{Albert2000}%
\bibitem{Jeong2001}%
  \BibitemOpen
  \bibfield{author}{%
  \bibinfo {author} {\bibfnamefont{H.}~\bibnamefont{Jeong}}, \bibinfo {author}
  {\bibfnamefont{S.}~\bibnamefont{Mason}}, \bibinfo {author}
  {\bibfnamefont{A.-L.}\ \bibnamefont{Barab\'{a}si}},\ and\ \bibinfo {author}
  {\bibfnamefont{Z.}~\bibnamefont{Oltvai}},\ }%
  \bibfield{journal}{%
  \bibinfo {journal} {Nature}\ }%
  \textbf{\bibinfo {volume} {411}},\ \bibinfo {pages} {41} (\bibinfo {month}
  {May}\ \bibinfo {year} {2001})%
  \bibAnnoteFile{NoStop}{Jeong2001}%
\bibitem{Holme2002}%
  \BibitemOpen
  \bibfield{author}{%
  \bibinfo {author} {\bibfnamefont{P.}~\bibnamefont{Holme}}, \bibinfo {author}
  {\bibfnamefont{B.}~\bibnamefont{Kim}}, \bibinfo {author}
  {\bibfnamefont{C.}~\bibnamefont{Yoon}},\ and\ \bibinfo {author}
  {\bibfnamefont{S.}~\bibnamefont{Han}},\ }%
  \bibfield{journal}{%
  \bibinfo {journal} {Physical Review E}\ }%
  \textbf{\bibinfo {volume} {65}},\ \bibinfo {pages} {056109} (\bibinfo {month}
  {May}\ \bibinfo {year} {2002})%
  \bibAnnoteFile{NoStop}{Holme2002}%
\bibitem{Albert2004}%
  \BibitemOpen
  \bibfield{author}{%
  \bibinfo {author} {\bibfnamefont{R.}~\bibnamefont{Albert}}, \bibinfo {author}
  {\bibfnamefont{I.}~\bibnamefont{Albert}},\ and\ \bibinfo {author}
  {\bibfnamefont{G.}~\bibnamefont{Nakarado}},\ }%
  \bibfield{journal}{%
  \bibinfo {journal} {Physical Review E}\ }%
  \textbf{\bibinfo {volume} {69}} (\bibinfo {month} {Feb.}\ \bibinfo {year}
  {2004})%
  \bibAnnoteFile{NoStop}{Albert2004}%
\bibitem{Crucitti2004}%
  \BibitemOpen
  \bibfield{author}{%
  \bibinfo {author} {\bibfnamefont{P.}~\bibnamefont{Crucitti}}, \bibinfo
  {author} {\bibfnamefont{V.}~\bibnamefont{Latora}},\ and\ \bibinfo {author}
  {\bibfnamefont{M.}~\bibnamefont{Marchiori}},\ }%
  \bibfield{journal}{%
  \bibinfo {journal} {Physical Review E}\ }%
  \textbf{\bibinfo {volume} {69}},\ \bibinfo {pages} {045104} (\bibinfo {month}
  {Apr.}\ \bibinfo {year} {2004})%
  \bibAnnoteFile{NoStop}{Crucitti2004}%
\bibitem{Estrada2006}%
  \BibitemOpen
  \bibfield{author}{%
  \bibinfo {author} {\bibfnamefont{E.}~\bibnamefont{Estrada}},\ }%
  \bibfield{journal}{%
  \bibinfo {journal} {The European Physical Journal B}\ }%
  \textbf{\bibinfo {volume} {52}},\ \bibinfo {pages} {563} (\bibinfo {month}
  {Aug.}\ \bibinfo {year} {2006})%
  \bibAnnoteFile{NoStop}{Estrada2006}%
\bibitem{Montoya2006}%
  \BibitemOpen
  \bibfield{author}{%
  \bibinfo {author} {\bibfnamefont{J.~M.}\ \bibnamefont{Montoya}}, \bibinfo
  {author} {\bibfnamefont{S.~L.}\ \bibnamefont{Pimm}},\ and\ \bibinfo {author}
  {\bibfnamefont{R.~V.}\ \bibnamefont{Sol\'{e}}},\ }%
  \bibfield{journal}{%
  \bibinfo {journal} {Nature}\ }%
  \textbf{\bibinfo {volume} {442}},\ \bibinfo {pages} {259} (\bibinfo {month}
  {Jul.}\ \bibinfo {year} {2006})%
  \bibAnnoteFile{NoStop}{Montoya2006}%
\bibitem{Liu2011}%
  \BibitemOpen
  \bibfield{author}{%
  \bibinfo {author} {\bibfnamefont{Y.-Y.}\ \bibnamefont{Liu}}, \bibinfo
  {author} {\bibfnamefont{J.-J.}\ \bibnamefont{Slotine}},\ and\ \bibinfo
  {author} {\bibfnamefont{A.-L.}\ \bibnamefont{Barab\'{a}si}},\ }%
  \bibfield{journal}{%
  \bibinfo {journal} {Nature}\ }%
  \textbf{\bibinfo {volume} {473}},\ \bibinfo {pages} {167} (\bibinfo {month}
  {May}\ \bibinfo {year} {2011})%
  \bibAnnoteFile{NoStop}{Liu2011}%
\bibitem{Albert2002}%
  \BibitemOpen
  \bibfield{author}{%
  \bibinfo {author} {\bibfnamefont{R.}~\bibnamefont{Albert}}\ and\ \bibinfo
  {author} {\bibfnamefont{A.-L.}\ \bibnamefont{Barab\'{a}si}},\ }%
  \bibfield{journal}{%
  \bibinfo {journal} {Reviews of Modern Physics}\ }%
  \textbf{\bibinfo {volume} {74}},\ \bibinfo {pages} {47} (\bibinfo {year}
  {2002})%
  \bibAnnoteFile{NoStop}{Albert2002}%
\bibitem{Watts1998}%
  \BibitemOpen
  \bibfield{author}{%
  \bibinfo {author} {\bibfnamefont{D.~J.}\ \bibnamefont{Watts}}\ and\ \bibinfo
  {author} {\bibfnamefont{S.~H.}\ \bibnamefont{Strogatz}},\ }%
  \bibfield{journal}{%
  \bibinfo {journal} {Nature}\ }%
  \textbf{\bibinfo {volume} {393}},\ \bibinfo {pages} {440} (\bibinfo {month}
  {Jun.}\ \bibinfo {year} {1998})%
  \bibAnnoteFile{NoStop}{Watts1998}%
\bibitem{Freeman1977}%
  \BibitemOpen
  \bibfield{author}{%
  \bibinfo {author} {\bibfnamefont{L.~C.}\ \bibnamefont{Freeman}},\ }%
  \bibfield{journal}{%
  \bibinfo {journal} {Sociometry}\ }%
  \textbf{\bibinfo {volume} {40}},\ \bibinfo {pages} {35} (\bibinfo {year}
  {1977})%
  \bibAnnoteFile{NoStop}{Freeman1977}%
\bibitem{Ferrari2004a}%
  \BibitemOpen
  \bibfield{author}{%
  \bibinfo {author} {\bibfnamefont{S.}~\bibnamefont{Ferrari}}\ and\ \bibinfo
  {author} {\bibfnamefont{F.}~\bibnamefont{Cribari-Neto}},\ }%
  \bibfield{journal}{%
  \bibinfo {journal} {Journal of Applied Statistics}\ }%
  \textbf{\bibinfo {volume} {31}},\ \bibinfo {pages} {799} (\bibinfo {month}
  {Aug.}\ \bibinfo {year} {2004})%
  \bibAnnoteFile{NoStop}{Ferrari2004a}%
\bibitem{Cribari2010}%
  \BibitemOpen
  \bibfield{author}{%
  \bibinfo {author} {\bibfnamefont{F.}~\bibnamefont{Cribari-Neto}}\ and\
  \bibinfo {author} {\bibfnamefont{A.}~\bibnamefont{Zeileis}},\ }%
  \bibfield{journal}{%
  \bibinfo {journal} {Journal of Statistical Software}\ }%
  \textbf{\bibinfo {volume} {34}} (\bibinfo {year} {2010})%
  \bibAnnoteFile{NoStop}{Cribari2010}%
\bibitem{R2012}%
  \BibitemOpen
  \bibfield{author}{%
  \bibinfo {author} {\bibnamefont{{R Core Team}}},\ }%
  \emph{\bibinfo {title} {R: A Language and Environment for Statistical
  Computing}},\ \bibinfo {organization} {R Foundation for Statistical
  Computing},\ \bibinfo {address} {Vienna, Austria} (\bibinfo {year} {2012})%
  \bibAnnoteFile{NoStop}{R2012}%
\bibitem{Huxham1996}%
  \BibitemOpen
  \bibfield{author}{%
  \bibinfo {author} {\bibfnamefont{M.}~\bibnamefont{Huxham}}, \bibinfo {author}
  {\bibfnamefont{S.}~\bibnamefont{Beaney}},\ and\ \bibinfo {author}
  {\bibfnamefont{D.}~\bibnamefont{Raffaelli}},\ }%
  \bibfield{journal}{%
  \bibinfo {journal} {Oikos}\ }%
  \textbf{\bibinfo {volume} {76}},\ \bibinfo {pages} {284} (\bibinfo {year}
  {1996})%
  \bibAnnoteFile{NoStop}{Huxham1996}%
\bibitem{Keseler2011}%
  \BibitemOpen
  \bibfield{author}{%
  \bibinfo {author} {\bibfnamefont{I.~M.}\ \bibnamefont{Keseler}}, \bibinfo
  {author} {\bibfnamefont{J.}~\bibnamefont{Collado-Vides}}, \bibinfo {author}
  {\bibfnamefont{A.}~\bibnamefont{Santos-Zavaleta}}, \bibinfo {author}
  {\bibfnamefont{M.}~\bibnamefont{Peralta-Gil}}, \bibinfo {author}
  {\bibfnamefont{S.}~\bibnamefont{Gama-Castro}}, \bibinfo {author}
  {\bibfnamefont{L.}~\bibnamefont{Mu\~{n}iz Rascado}}, \bibinfo {author}
  {\bibfnamefont{C.}~\bibnamefont{Bonavides-Martinez}}, \bibinfo {author}
  {\bibfnamefont{S.}~\bibnamefont{Paley}}, \bibinfo {author}
  {\bibfnamefont{M.}~\bibnamefont{Krummenacker}}, \bibinfo {author}
  {\bibfnamefont{T.}~\bibnamefont{Altman}}, \bibinfo {author}
  {\bibfnamefont{P.}~\bibnamefont{Kaipa}}, \bibinfo {author}
  {\bibfnamefont{A.}~\bibnamefont{Spaulding}}, \bibinfo {author}
  {\bibfnamefont{J.}~\bibnamefont{Pacheco}}, \bibinfo {author}
  {\bibfnamefont{M.}~\bibnamefont{Latendresse}}, \bibinfo {author}
  {\bibfnamefont{C.}~\bibnamefont{Fulcher}}, \bibinfo {author}
  {\bibfnamefont{M.}~\bibnamefont{Sarker}}, \bibinfo {author}
  {\bibfnamefont{A.~G.}\ \bibnamefont{Shearer}}, \bibinfo {author}
  {\bibfnamefont{A.}~\bibnamefont{Mackie}}, \bibinfo {author}
  {\bibfnamefont{I.}~\bibnamefont{Paulsen}}, \bibinfo {author}
  {\bibfnamefont{R.~P.}\ \bibnamefont{Gunsalus}},\ and\ \bibinfo {author}
  {\bibfnamefont{P.~D.}\ \bibnamefont{Karp}},\ }%
  \bibfield{journal}{%
  \bibinfo {journal} {Nucleic acids research}\ }%
  \textbf{\bibinfo {volume} {39}},\ \bibinfo {pages} {D583} (\bibinfo {month}
  {Jan.}\ \bibinfo {year} {2011})%
  \bibAnnoteFile{NoStop}{Keseler2011}%
\bibitem{Shaikh2007}%
  \BibitemOpen
  \bibfield{author}{%
  \bibinfo {author} {\bibfnamefont{M.~A.}\ \bibnamefont{Shaikh}}, \bibinfo
  {author} {\bibfnamefont{J.}~\bibnamefont{Wang}}, \bibinfo {author}
  {\bibfnamefont{Z.}~\bibnamefont{Yang}},\ and\ \bibinfo {author}
  {\bibfnamefont{Y.}~\bibnamefont{Song}},\ }%
  in\ \emph{\bibinfo {booktitle} {Advanced Data Mining and Applications}},\
  Vol.\ \bibinfo {volume} {4362/2007}\ (\bibinfo {publisher} {Springer
  Berlin},\ \bibinfo {year} {2007})\ pp.\ \bibinfo {pages} {570--577}%
  \bibAnnoteFile{NoStop}{Shaikh2007}%
\end{thebibliography}%

\end{document}